\documentclass[modern]{aastex61}
\usepackage{natbib}
\received{November 3, 2017}
\revised{November 8, 2017}
\accepted{November 8, 2017}
\submitjournal{Astrophysical Journal Letters}

\shorttitle{Planetary system formation from 1I/2017 U1 (`Oumuamua)}
\shortauthors{Trilling et al.}

\begin{document}

\title{Implications for planetary system formation
from interstellar object 1I/2017 U1 (`Oumuamua)}

\correspondingauthor{David E. Trilling}
\email{david.trilling@nau.edu}

\author{David E. Trilling}
\affil{Department of Physics and Astronomy \\
Northern Arizona University \\
PO Box 6010 \\
Flagstaff, AZ 86011}

\author{Tyler Robinson}
\affil{Department of Physics and Astronomy \\
Northern Arizona University \\
PO Box 6010 \\
Flagstaff, AZ 86011}

\author{Alissa Roegge}
\affil{Department of Physics and Astronomy \\
Northern Arizona University \\
PO Box 6010 \\
Flagstaff, AZ 86011}

\author[0000-0001-7335-1715]{Colin Orion Chandler}
\affil{Department of Physics and Astronomy \\
Northern Arizona University \\
PO Box 6010 \\
Flagstaff, AZ 86011}

\author{Nathan Smith}
\affil{Department of Physics and Astronomy \\
Northern Arizona University \\
PO Box 6010 \\
Flagstaff, AZ 86011}

\author{Mark Loeffler}
\affil{Department of Physics and Astronomy \\
Northern Arizona University \\
PO Box 6010 \\
Flagstaff, AZ 86011}

\author{Chad Trujillo}
\affil{Department of Physics and Astronomy \\
Northern Arizona University \\
PO Box 6010 \\
Flagstaff, AZ 86011}

\author{Samuel Navarro-Meza}
\affil{Department of Physics and Astronomy \\
Northern Arizona University \\
PO Box 6010 \\
Flagstaff, AZ 86011}
\affil{Instituto de Astronom\'ia \\
Universidad Nacional Aut\'onoma de M\'exico \\
Ensendada B.C, 22860 \\
M\'exico}

\author{Lori M. Glaspie}
\affil{Department of Physics and Astronomy \\
Northern Arizona University \\
PO Box 6010 \\
Flagstaff, AZ 86011}



\begin{abstract}
The recently discovered minor body
1I/2017 U1 (`Oumuamua)
is the first known object
in our Solar System that is not bound by the 
Sun's gravity. Its
hyperbolic orbit (eccentricity greater than unity)
strongly suggests 
that it originated outside our Solar
System; its red color is consistent
with substantial space weathering experienced over
a long interstellar journey.
We carry out an simple calculation
of the probability of detecting such an object.
We find that the observed detection rate
of 1I-like objects can be satisfied
if the average mass of ejected material
from nearby stars during the process of
planetary formation is $\sim$20~Earth masses,
similar to the expected value for our Solar System.
The current detection rate of such interstellar
interlopers is estimated to be~0.2/year,
and the expected number of detections over
the past few years is almost exactly one.
When
the Large Synoptic Survey Telescope
begins its wide, fast, deep all-sky survey
the detection rate will increase to~1/year.
Those expected detections will
provide
further constraints on nearby planetary
system formation through a better estimate of
the number and properties of interstellar objects.
\end{abstract}




\section{Introduction} \label{intro}

On October 18, 2017 
the minor body C/2017-UT (PANSTARRS) that would become 
known as A/2017~U1 
and later 1I/2017 U1 (`Oumuamua) --- hereafter,
1I ---
was discovered
by the the Panoramic Survey Telescope And Rapid Response System (Pan-STARRS) 
survey \citep{Chambers:2016vk}. In a matter of days it 
became clear that this object was on an unbound
(hyperbolic) trajectory, 
with eccentricity $e\approx 1.2$ \citep{Williams:2017wa}. 
In addition to its high eccentricity, 1I's orbit 
is inclined to the Solar System plane at an angle $i\approx 123^\circ$,
so it is very unlikely to have encountered any
massive objects within our Solar System (other than the Sun).
With no gravitational perturbations to explain its
anomalously high eccentricity, 
the most likely explanation is that this object
originated outside our Solar System and happened
to pass close to the Earth during its journey
through interstellar space. This implies that 1I 
formed in another planetary system and was
ejected, presumably through dynamical interactions
in its natal planetary system.
Consequently, we refer to 1I and similar bodies as 
``ejectoids.''

The theoretical existence of ejectoids has been long proposed
\citep[e.g.,][]{1976Icar...27..123S},
with various authors using non-detections to place upper limits on the population
\citep{1989ApJ...346L.105M,1993A&A...275..298S,2017AJ....153..133E}.
\citet{2003EM&P...92..465J}
and 
\citet{2005ApJ...635.1348F}
suggested that PanSTARRS could detect an interstellar object, if the number density
was great enough. 
In the two weeks since the discovery of 1I and
its identification as an interstellar object, we have learned
about its optical spectrum \citep{Masiero:2017vr,qz},
its lightcurve \citep{knight},
its orbit \citep{2017arXiv171100445D}, and potential
origin scenarios
\citep{2017arXiv171011364M,gaidos,laughlin}.

%

Here we use a simple calculation to 
estimate the probability of detecting such an object
and
explore the implications for the prevalence
and properties of planetary systems that are implied
by the existence and properties of 1I.


\section{Ejectoid Encounters and Ejected Mass Constraints}
 
Gravitational microlensing results have shown that, on average, every star in the Milky Way is accompanied
by at least one bound planet \citep{Cassan:2012in}, which implies that planet formation is a near-universal
process.  We assume here that the formation of a typical planetary system results in a mass $m$ of ejectoids
with 
a typical size of $r_{\rm{e}}$ and mass density $\rho_{\rm{e}}$. The number of ejectoids
per star is therefore

\begin{equation}
  \frac{m}{\frac{4}{3}\pi r_{\rm{e}}^3  \rho_{\rm{e}}} \ .
\end{equation}

We can write the number density of stars (in, e.g.,~stars per cubic parsec)
$n_{\rm{s}}$ 
in terms of a characteristic stellar spacing, $R$, as

\begin{equation}
  n_{\rm{s}} = \frac{1}{\frac{4}{3}\pi R^3} \ .
\end{equation}

This enables us to express the number density of ejectoids (number/volume) as

\begin{equation}
  n_{\rm{e}} = \left( \frac{m}{\frac{4}{3}\pi r_{\rm{e}}^3  \rho_{\rm{e}}} \right) \left( \frac{1}{\frac{4}{3}\pi R^3} \right) \ .
\label{numberdensity}
\end{equation}
 
We next wish to determine the number of ejectoids that we
expect to have encountered.  We assume that telescopic searches for 1I-like objects have swept
out a cylindrical volume of interstellar space given by

\begin{equation}
  V_{\rm{obs}} = \pi r_{\rm{obs}}^2 \Delta v_{\odot} \Delta t
\end{equation}

where $r_{\rm{obs}}$ is the geocentric distance out to which we are sensitive to 1I-like objects,
$\Delta v_{\odot}$ is the Sun's velocity through the solar
neighborhood (and the presumed cloud of ejectoids), and $\Delta t$ is the time interval
over which observational surveys have been capable of discovering 1I-like objects.  Using this
volume and the number density of ejectoids, we find that the number of detections of 1I-like objects
is

\begin{equation}
  N = n_{\rm{e}} V_{\rm{obs}} = \left( \frac{m}{\frac{4}{3}\pi r_{\rm{e}}^3  \rho_{\rm{e}}} \right) \left( \frac{1}{\frac{4}{3}\pi R^3} \right) \left(\pi r_{\rm{obs}}^2 \Delta v_{\odot} \Delta t \right)
\end{equation}

which simplifies to

\begin{equation}
  N = \frac{9}{16\pi} \frac{ m r_{\rm{obs}}^2 \Delta v_{\odot} \Delta t}{\rho_{\rm{e}}r_{\rm{e}}^3 R^3} \ .
\label{simplified}
\end{equation}
 
We can now use 1I to constrain the characteristic mass in ejectoids for a forming
planetary system.  With $N = 1$, we can re-arrange Equation~\ref{simplified} to write that

\begin{equation}
  m = \frac{16\pi}{9} \frac{\rho_{\rm{e}}r_{\rm{e}}^3 R^3}{r_{\rm{obs}}^2 \Delta v_{\odot} \Delta t} \ .
\label{masseqn}
\end{equation}

The absolute magnitude of 1I is given in the 
Minor Planet Center catalog as~22.1 
(as of 2 November 2017), which 
corresponds 
to a radius $r_e$ 
of
around 100~meters,
assuming a moderate-to-dark albedo.

The range of densities for cometary and asteroidal 
material that may be relevant is around
500~to 3000~kg/m$^3$
\citep{2007Icar..190..357R,2006Icar..180..224D,2007Icar..187..306D,2012P&SS...73...98C}.
While 1I's highly eccentric orbit would be typically associated with a comet, and theoretical predictions suggest that most ejectoids should be more comet-like \citep{raymond},
there are no signs of activity from 1I \citep{knight,qz}.
Thus, here, we assume that 1I may be more asteroidal than cometary and adopt a density of 2000 kg/m$^3$.  

There are 357~stars within 10~parsecs
of the Sun\footnote{{\tt www.recons.org}}. This gives an average distance between
adjacent stars $R$
of
1.4~pc.
The average discovery distance of Near Earth Objects
in the Minor Planet Center $r_{\mathrm{obs}}$ is 0.3~au.

The velocity of the Sun relative to nearby
stars is around 20~km/sec \citep{2010MNRAS.403.1829S}.
1I has been found\footnote{{\tt https://projectpluto.com/temp/2017u1.htm}}
to have an interstellar speed (velocity at
infinity) of 26~km/sec, while
\citet{2017arXiv171011364M}
reports that an object entering the
Solar System with median velocity of the local
stellar population would have a speed of around 22.5~km/sec.
The fact that the Sun's relative velocity is
comparable to the velocity of 1I 
confirms our assumption that the population of interstellar
ejectoids has zero mean velocity (due to the 
fact that both the
source planetary systems and ejection trajectories
are assumed to be isotropic).
Based on these three estimates
we set $\Delta v_\odot$ to be 25~km/sec.

%
%
%

In recent years improvements in detector size and
field of view at the Catalina Sky Survey and
Pan-STARRS (the two major NEO 
surveys) have enhanced the ability to detect
1I-like objects, so we estimate $\Delta t$ to be
5~years.


Our characteristic values, when inserted
in Equation~\ref{masseqn},  yield a typical mass in ejected 1I-like objects of $10^{26}$~kg, or
20~$M_{\oplus}$.  This is in remarkably good agreement with values derived 
for mass loss during the formation of our Solar System.
For example, \cite{Weidenschilling:1977kq} and \cite{Bottke:2005jx} derive 1--5~$M_{\oplus}$ of material lost
from the asteroid belt.  \cite{Kuiper:1951hw}, \citet{1999AJ....118.1101K}, and \cite{Morbidelli:2005tx} find 12--30~$M_{\oplus}$ lost from
the Kuiper Belt. Together, these imply a total mass lost 
from our Solar System of close to 20~$M_\oplus$.

Identifying the characteristic size and density of ejectoids as being the most uncertain
terms in our analysis, and inserting the parameter values adopted above, we can write

\begin{equation}
  m = 20~M_{\oplus} \left( \frac{\rho_{e}}{2000~\rm{kg~m^{-3}}}\right) \left( \frac{r_{e}}{100~\rm{m}}\right)^3 \ .
\label{scaling}
\end{equation}

\begin{figure}[ht!]
\plotone{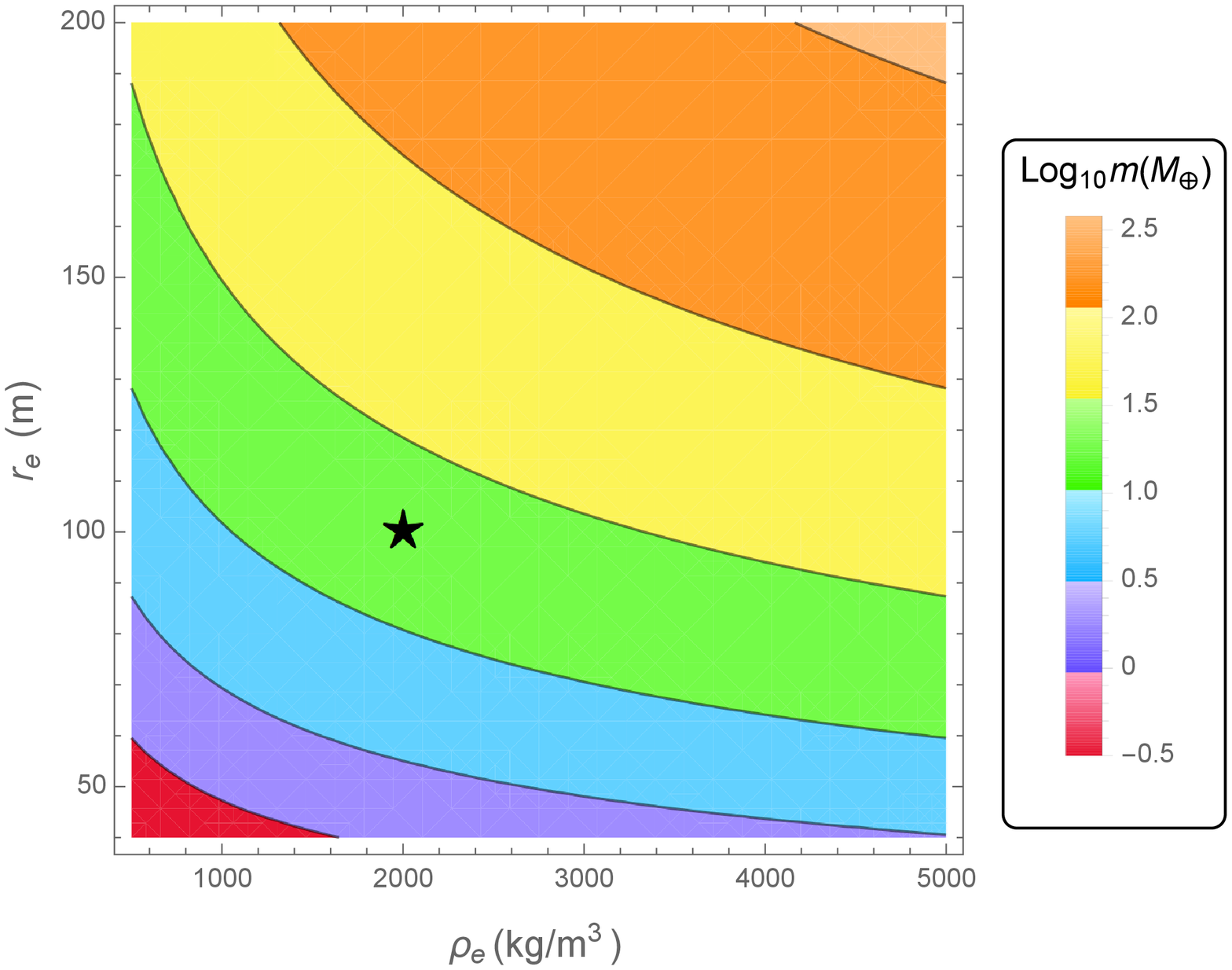}
\caption{Characteristic ejection mass (colors and contours, in 
Earth masses)
required to produce the observed 
rate of 1I-like objects as a function of the primary unknown parameters
in our analysis:
1I's density ($\rho_{\mathrm e}$) and radius ($r_{\mathrm e}$). For nominal values of 
2000~kg/m$^3$ and 100~m, respectively, the required
average ejection mass per star in the solar neighborhood
is 20~$M_\oplus$ (black star), remarkably close to various calculations
of the mass lost from our Solar System during the 
era of planet formation.
Our upcoming observations of 1I with the Spitzer 
Space Telescope in late November, 2017, should place
a constraint on 1I's radius.}
\label{contour}
\end{figure}

This relationship is shown in Figure~\ref{contour}.
The plausible range of ejection masses is
roughly 1--100~$M_\oplus$.
We note that in principle
the radius of 1I, which we take to be characteristic
of ejectoids,
will be determined through our
forthcoming thermal infrared observations
of 1I with the Spitzer Space Telescope
(observations scheduled for late November, 2017).
There is no obvious way to constrain the density
of 1I.
 
Alternatively, if we adopt the $\sim$20~$M_{\oplus}$ in ejectoids lost during the formation of the Solar System
as a characteristic number, we can use Equation~\ref{simplified} to derive the 
number of 1I-like objects expected over five years. We find
that N is very close to unity --- exactly matching
the observations.
The detection rate ($N/\Delta t$) is therefore 0.2~1I-like objects per year,
or one 1I-like
object every five years.
The number density of ejectoids is, from Equation~\ref{numberdensity}, 
around 0.1/pc$^3$.

%
%

\section{Caveats and uncertainties}

We do not claim that this is the only mechanism
for producing 1I-like objects or delivering
such objects to the detectable space near
the Earth, as the above calculation admittedly contains a number of assumptions.
However, this does give a plausible
explanation for 1I that in turn has
several interesting implications that are 
discussed below.

The radius and density of 1I are
unknown, though the values above are unlikely
to be in error by more than a factor of two.
Similarly, the geometry arguments (average stellar distance,
solar velocity, observational distance)
are likely within a factor of two,
while the 
time interval is approximately correct.
We ignore gravitational focusing here.
The largest overall uncertainty is simply the
unknown statistical likelihood of detecting
this ejectoid and our extrapolation
from a single object. 1I may be part of a constant
stream of interstellar objects moving through
our Solar System (as implied here), or a very
unlikely occurrence, in which case the arguments
made here are less applicable.

In addition, the true population of ejected extrasolar material
must follow some size distribution, and will not consist of only 100~m 1I-like objects.
Small objects are presumably more numerous in any 
planet formation scenario, but larger objects will
be preferentially detected by our surveys.
We must simply take 1I to be representative.

In the above calculation we have assumed 
that a steady state population
of 1I-like objects is ejected from all planetary
systems.
However, we might instead assume that the
majority of ejectoids are produced during
the earliest phases of
planetary system formation, in a single
pulse of material. The nearest
star formation regions are some 
100~parsec away, with typical ages 
of 1--10~Myr 
\citep{2009ApJ...700.1502A,2011ApJ...732...24C,2017AJ....154..134E}.
If we take the escape velocity from those systems to 
be on the order of 10~km/sec then material
from one of these nearby star formation
regions would reach the Earth
in a few million years, and we would therefore
be moving through a cloud of ejected 1I-like
objects. However, the rest of the assumptions
still apply, and the expected value does
not change significantly.
A more complicated model (perhaps not warranted, given this
single detection) could 
account for the 
total number of stars contained in the
Milky Way, integrated over its history, as
even stars that no longer exist could have
contributed ejectoids to an interstellar
stream of material.

Finally, we note that the above calculation implies 
a typical ejected mass of 20~$M_\oplus$, but that 
need not imply that every stellar or planetary system
ejects mass, or that amount of material.
For example, while planet-planet scattering among
gas giant planets is likely to produce a pulse of ejected planetesimals
\citep{raymond,marzari},
and while systems with gas giants on stable orbits
can eject planetesimals on longer timescales
\citep{raymond,barclay},
systems without gas giants
rarely eject planetesimals because in that case escape velocity
can not readily be achieved by planetesimals.
Thus, if the fraction of nearby stars with gas giants
is (for example) 50\% then the average mass ejected
by those systems must be a factor of~$\sim$2 greater
than our nominal value in order to produce the necessary
interstellar density of ejectoids.


\section{Implications}

The 520--950~nm reflectance spectrum of 1I, albeit noisy, indicates no absorption features and a red
spectral slope \citep{Masiero:2017vr,qz}, making compositional 
characterization difficult. However, we note that a red spectral slope over this 
range is not unexpected and is characteristic of many primitive objects in our Solar System,
\citep{Cruikshank:1998iy,1998AJ....115.1667J,Bus:2002ia,2010AJ....139.1394S,2016Icar..268..340C}.
Whether this slope is something intrinsic to the bulk properties of 1I or a consequence of its surface being altered via energetic processing
is unclear.
Given the presence of cosmic rays in the interstellar medium and other forms of ionizing radiation 
that would have been present in 1I's natal stellar environment, 
1I's red spectral slope is entirely consistent with 
formation elsewhere and a 
long journey to our Solar System.
This implies that the surface of 1I may have
different properties
than its bulk material.



1I's trajectory makes it very unlikely that it experienced
a gravitational encounter with any of the proposed as-yet
unknown planets in the outermost part of our Solar
System \citep{2014Natur.507..471T,2016ApJ...824L..23B,2017AJ....154...62V}.
Another possibility is that 1I was a member of our Solar 
System's Oort Cloud and was perturbed inbound onto an
unbound orbit by a passing star.
We do not comment on these scenarios; in this work we have 
assumed that 1I is an interstellar interloper that originated
in a different plantary system.

The Large Synoptic Survey
Telescope (LSST; \citealt{Ivezic:2008ub}) will commence its ten year 
all-sky survey in 2022. One of the driving science
cases for LSST is the detection of moving 
objects. Most moving objects detected
by LSST will be ``unremarkable'' asteroids
in the main belt, but this very large 
and deep survey (20,000~deg$^2$ surveyed
to $r$~magnitude~24.5 repeatedly over ten years)
naturally has the possibility to discover
unusual objects in all areas of astrophysics.
Several authors have
studied the detectability of
interstellar interlopers in LSST data
\citep{2009ApJ...704..733M,2016ApJ...825...51C,2017AJ....153..133E}.

LSST
will help
constrain nearby planetary system formation
by measuring the number of 1I-like objects.
LSST will be sensitive to fainter and
therefore smaller and/or more distant
objects, and is therefore likely to have
a greater detection rate than the current
rate.
The possibility of LSST detecting interstellar comets 
increases order(s) of magnitude when considering cometary outbursting
\citep{2016ApJ...825...51C},
which makes objects brighter, although 1I has shown no signs
of activity so far in its observational record.

As described above, with our current detection
sensitivities the detection rate for
1I-like objects is~0.2/year.
The LSST detection limit will be around
three magnitudes deeper than Pan-STARRS' typical
limiting magnitude of V$\sim$21.5; this
translates to a factor of three smaller in size.
The only measured size distribution
in this size range in our Solar System 
is
the Near Earth Object population;
a factor of three in 
size corresponds roughly to a factor of five
in number of objects \citep{2017AJ....154..170T}.
Thus, the expected detection rate of 
interstellar objects for LSST is 
around~1/year.
The LSST discovery rate of ejectoids will
help us constrain the frequency and properties
of planetary system formation in our nearby galaxy.

\acknowledgments

We thank an anonymous referee for a very prompt
and thoughtful review that has improved this 
paper.
This research has made use of data and/or services provided by the International Astronomical Union's Minor Planet Center.
We have used information from the RECONS program, retrieved from
{\tt www.recons.org} . 
The discussions that led to this paper began at
an NAU Astrocookies gathering.

\end{document}